\documentclass{article}
\usepackage{amsmath,amssymb,amsthm, amsfonts, mathrsfs}
\usepackage{calligra,indentfirst,epsfig, enumitem, lettrine}
\calligra
\usepackage{cite}
\usepackage[dvipsnames]{xcolor}

\theoremstyle{remark}

\theoremstyle{definition}

\allowdisplaybreaks
\numberwithin{equation}{section}

\begin{document}

\title{Some remarks on the results derived by\\ Ramy Takieldin and Patrick Sol\'e (2025)}
\author{Varsha Chauhan{\footnote{Email address: vchauhan@maitreyi.du.ac.in}} ~and Anuradha Sharma{\footnote{Email address: anuradha@iiitd.ac.in} }\\$^\ast$Department of Mathematics, Maitreyi College, New Delhi 110021, India\\$^\dagger$Department of Mathematics, IIIT-Delhi, New Delhi 110020, India}
\date{}
\maketitle
\begin{abstract}
The purpose of this note is to rectify a typographical error in the statements of  Theorems 5.5 and 5.6 of Sharma, Chauhan and Singh \cite{multi} and further analyze and discuss the significance of the results derived in Takieldin and Sol\'e \cite{ramy}. In our opinion, several claims made by the authors in \cite{ramy} are either factually incorrect or  lack adequate substantiation, which may confuse the readers about the contributions of \cite{multi2, multi}. Our remarks on the work \cite{ramy} intend to provide the clarity and inform about the  true contributions and findings of our research. 
\end{abstract}
\hrulefill \vspace{2mm}\\
In \cite{ramy}, the authors begin by presenting counterexamples to Theorems 5.5 and 5.6 of \cite{multi}. Additionally, they made an unsubstantiated claim that ``these examples emphasize that imposing a condition on the dimension of the multi-twisted (MT) code or its dual to guarantee being LCD is impossible. For instance, the MT code given in Example 1 has dimension greater than $\min\{m_i\},$ while its dual has dimension smaller than $\min\{m_i\},$ and that in Example 2 has dimension equal to $\min\{m_i\}$, and all of which are not LCD " (see page 4, lines 2 - 5 of the last paragraph). 

The purpose of this note is to address a typographical error in the statements of Theorems 5.5 and 5.6 of \cite{multi}, which can be corrected by replacing ``$\Lambda$-multi-twisted code" with ``$[\Lambda, \Lambda']$-multi-twisted code". The proofs of Theorems 5.5 and 5.6 of \cite{multi} are correct. Moreover, Corollary 5.1 of \cite{multi} is correct and follows directly from Theorem 5.2 of \cite{multi} and Corollary 2.6 of Liu and Liu \cite{Liu}.
This correction also resolves the issue raised by the authors of \cite{ramy} in Remark 3.1(b) of \cite{multi2}, where it is stated that the results derived in Section 5 of \cite{multi} can be similarly extended when the block lengths $m_i$\textquotesingle s
 are not necessarily coprime to $q,$ particularly the BCH type distance bound obtained in Theorem 5.3 of \cite{multi}.

 Another purpose of this note is to analyze and discuss the significance of the results derived in \cite{ramy}. For this, we first state Theorem 7 of \cite{ramy}. \\
\noindent\textbf{Theorem 7 of \cite{ramy}.} {\it Let $\Lambda=\left(\lambda_1,\lambda_2,\ldots,\lambda_\ell\right)$, where $\lambda_i$ is a non-zero element of $\mathbb{F}_q$ for $1\le i\le \ell$. Let $\mathcal{C}$ (or $\mathcal{C}^\perp$) be a $\Lambda$-MT code of block lengths $\left(m_1,m_2,\ldots,m_\ell\right)$ generated by $\left\{\mathbf{g}_1, \mathbf{g}_2, \ldots, \mathbf{g}_\rho\right\}$. For $1\le i\le \ell$, let $g_i\left(x\right)=\mathrm{gcd}\left\{\left(x^{m_i}-\lambda_i\right),\pi_i\left(\mathbf{g}_1\right),\pi_i\left(\mathbf{g}_2\right),\ldots,\pi_i\left(\mathbf{g}_\rho\right)\right\}$. If 
$$\frac{x^{m_1}-\lambda_1}{g_1\left(x\right)}, \frac{x^{m_2}-\lambda_2}{g_2\left(x\right)}, \ldots, \frac{x^{m_\ell}-\lambda_\ell}{g_\ell\left(x\right)}$$ are pairwise coprime polynomials in $\mathbb{F}_q[x]$, then $\mathcal{C}$ and $\mathcal{C}^\perp$ are direct sums of $\ell$ constacyclic codes, in particular
\begin{equation*}
\begin{split}
\mathcal{C}&=\bigoplus_{i=1}^\ell \pi_i\left(\mathcal{C}\right)\\
\mathcal{C}^\perp&=\bigoplus_{i=1}^\ell \pi_i\left(\mathcal{C}\right)^\perp=\bigoplus_{i=1}^\ell \pi_i\left(\mathcal{C}^\perp\right).
\end{split}
\end{equation*}}

\noindent\textbf{Remark 1.} {\it The above theorem implies that when $\frac{x^{m_1}-\lambda_1}{g_1\left(x\right)}, \frac{x^{m_2}-\lambda_2}{g_2\left(x\right)}, \ldots, \frac{x^{m_\ell}-\lambda_\ell}{g_\ell\left(x\right)}$ are pairwise coprime polynomials in $\mathbb{F}_q[x]$,  the $\Lambda$-multi-twisted code $\mathcal{C}$ (as defined in the above theorem) is a direct sum of the codes $\pi_1(\mathcal{C}),\pi_2(\mathcal{C}),\ldots, \pi_\ell(\mathcal{C}),$ where $\pi_i(\mathcal{C})$ is a $\lambda_i$-constacyclic code  of length $m_i$ over $\mathbb{F}_q$ for $1 \leq i \leq \ell.$ Further,  the Euclidean  dual code $\mathcal{C}^{\perp}$ is a direct sum of the dual codes $\pi_1(\mathcal{C})^{\perp},\pi_2(\mathcal{C})^{\perp},\ldots, \pi_\ell(\mathcal{C})^{\perp},$ where $\pi_i(\mathcal{C})^{\perp}$ is a $\lambda_i^{-1}$-constacyclic code of length $m_i$ over $\mathbb{F}_q$ for $1 \leq i \leq \ell.$ From this, it follows that if $G_i$ is a generator matrix of the $\lambda_i$-constacyclic code $\pi_i(\mathcal{C})$ of length $m_i$ over $\mathbb{F}_q,$ then the $\Lambda$-multi-twisted code $\mathcal{C}=\bigoplus\limits_{i=1}^\ell \pi_i\left(\mathcal{C}\right)$ of length $n=m_1+m_2+\cdots+m_{\ell}$ and block lengths $\left(m_1,m_2,\ldots,m_\ell\right)$ over $\mathbb{F}_q$ has a generator matrix $$G=\begin{bmatrix} G_1 & 0 & 0& \cdots & 0\\0& G_2 & 0 & \cdots & 0 \\
\cdots & \cdots & \cdots & \cdots & \cdots\\\cdots & \cdots & \cdots & \cdots & \cdots\\ 0 & 0 & 0& \cdots & G_{\ell}
    \end{bmatrix}.$$ This further implies that \begin{equation}\label{D}d(\mathcal{C})=\min \{d(\pi_1(\mathcal{C})), d(\pi_2(\mathcal{C})),\ldots, d(\pi_\ell(\mathcal{C}))\},\end{equation} where $d(\cdot)$ denotes the Hamming distance function. Thus, the Hamming distance (and consequently, the error-correction capability) of the $\Lambda$-multi-twisted code $\mathcal{C}$ cannot exceed the Hamming distances (and error-correction capabilities) of any of the constacyclic codes $\pi_i(\mathcal{C})$ for $1 \leq i \leq \ell$. Furthermore, the code $\mathcal{C}$ has length $n = m_1 + m_2 + \cdots + m_{\ell}$, while the constacyclic code $\pi_i(\mathcal{C})$ has length $m_i$ for $1 \leq i \leq \ell$. Consequently, the class of $\Lambda$-multi-twisted codes examined in Theorem 7 of \cite{ramy} lacks significant relevance from a coding theory perspective. Notably, not all multi-twisted codes over finite fields and their dual codes are direct sums of constacyclic codes. Thus, the majority of  results derived in \cite{multi2} and Section 5 of \cite{multi} focus on multi-twisted codes and their special subclasses, which are not necessarily direct sums of constacyclic codes.}

 In Theorem 5.2 of \cite{multi}, we obtained  a generating set of the Euclidean dual code of a $\Lambda$-multi-twisted code of length $n=m_1+m_2+\ldots+m_{\ell}$ and block lengths $\left(m_1,m_2,\ldots,m_\ell\right)$ over $\mathbb{F}_q$ when the polynomials $x^{m_1}-\lambda_1,x^{m_2}-\lambda_2,\ldots, x^{m_\ell}-\lambda_\ell$ are pairwise coprime in $\mathbb{F}_q[x].$ Later, in Theorem 3.1  of \cite{multi2} (\textit{resp.} Theorem 3.2 of \cite{multi2}), we showed that each $\Lambda$-multi-twisted code of length $n=m_1+m_2+\ldots+m_{\ell}$ and block lengths $\left(m_1,m_2,\ldots,m_\ell\right)$ over $\mathbb{F}_q$ has a normalized generating set (\textit{resp.} unique nice normalized generating set), and further, in Theorem 3.3 of \cite{multi2}, we  explicitly obtained a generating set of its dual code in terms of the normalized generating set of the $\Lambda$-multi-twisted code of length $n=m_1+m_2+\ldots+m_{\ell}$ and block lengths $\left(m_1,m_2,\ldots,m_\ell\right)$ over $\mathbb{F}_q$, where  the polynomials $x^{m_1}-\lambda_1,x^{m_2}-\lambda_2,\ldots, x^{m_\ell}-\lambda_\ell$ are not necessarily pairwise coprime in $\mathbb{F}_q[x]$ and $m_1, m_2,\ldots,m_{\ell}$ are arbitrary positive integers, not necessarily coprime to $q.$  However, in \cite{ramy}, the authors only discussed Theorem 5.2 of \cite{multi} and did not discuss Theorem 3.3 of \cite{multi2}, thereby providing  an incomplete and inaccurate representation of the existing results in this direction.
 
We next state Corollary 8 of \cite{ramy}.

\noindent\textbf{Corollary 8 of \cite{ramy}.} {\it 
Let $\Lambda=\left(\lambda_1,\lambda_2,\ldots,\lambda_\ell\right)$, where $0\ne\lambda_i\ne\lambda_i^{-1}$ for $1\le i\le \ell$. Let $\mathcal{C}$ (or $\mathcal{C}^\perp$) be a $\Lambda$-MT code of block lengths $\left(m_1,m_2,\ldots,m_\ell\right)$ generated by $\left\{\mathbf{g}_1, \mathbf{g}_2, \ldots, \mathbf{g}_\rho\right\}$. For $1\le i\le \ell$, let $g_i\left(x\right)=\mathrm{gcd}\left\{\left(x^{m_i}-\lambda_i\right),\pi_i\left(\mathbf{g}_1\right),\pi_i\left(\mathbf{g}_2\right), \ldots,\pi_i\left(\mathbf{g}_\rho\right)\right\}$. Assume 
$$\frac{x^{m_1}-\lambda_1}{g_1\left(x\right)}, \frac{x^{m_2}-\lambda_2}{g_2\left(x\right)}, \ldots, \frac{x^{m_\ell}-\lambda_\ell}{g_\ell\left(x\right)}$$ are pairwise coprime polynomials in $\mathbb{F}_q[x]$, then $\mathcal{C}$ is LCD.}

\noindent \textbf{Remark 2.} {\it By Theorem  7 of \cite{ramy}, the code $\mathcal{C}$ (as defined in Theorem 7 of \cite{ramy}) is LCD if and only if  $\pi_i(\mathcal{C})$ is LCD for $1 \leq i \leq \ell.$ By applying Corollary 2.6 of Liu and Liu \cite{Liu}, Corollary 8 of \cite{ramy} follows immediately. }

Theorem 9 of \cite{ramy} is as stated below.

\vspace{2mm}\noindent \textbf{Theorem 9 of \cite{ramy}.} {\it 
Let $\Lambda=\left(\lambda_1,\lambda_2,\ldots,\lambda_\ell\right)$, where $\lambda_i$ is a non-zero element of $\mathbb{F}_q$ for $1\le i\le \ell$. Let $\mathcal{C}$ (or $\mathcal{C}^\perp$) be a $\Lambda$-MT code of block lengths $\left(m_1,m_2,\ldots,m_\ell\right)$ generated by $\left\{\mathbf{g}_1, \mathbf{g}_2, \ldots, \mathbf{g}_\rho\right\}$. For $1\le i\le \ell$, let $g_i\left(x\right)=\mathrm{gcd}\left\{\left(x^{m_i}-\lambda_i\right),\pi_i\left(\mathbf{g}_1\right),\pi_i\left(\mathbf{g}_2\right), \ldots,\pi_i\left(\mathbf{g}_\rho\right)\right\}$. If 
$$\frac{x^{m_1}-\lambda_1}{g_1\left(x\right)}, \frac{x^{m_2}-\lambda_2}{g_2\left(x\right)}, \ldots, \frac{x^{m_\ell}-\lambda_\ell}{g_\ell\left(x\right)}$$ are pairwise coprime polynomials in $\mathbb{F}_q[x]$, then $\mathcal{C}$ is LCD if and only if $g_i\left(x\right)$ is self-reciprocal coprime to $\frac{x^{m_i}-\lambda_i}{g_i\left(x\right)}$ for every $i$ with $\lambda_i^2=1$.} 
\\ \noindent \textbf{Remark 3.} {\it  Note that Theorem 9 of \cite{ramy} is a straightforward consequence of Theorem 7 of \cite{ramy} and an easy extension of the main theorem derived in Yang and Massey \cite{massey} to $\lambda$-constacyclic codes over $\mathbb{F}_q,$ where $\lambda =\lambda^{-1}$.}

\noindent \textbf{Remark 4.} {\it When $\frac{x^{m_1}-\lambda_1}{g_1\left(x\right)}, \frac{x^{m_2}-\lambda_2}{g_2\left(x\right)}, \ldots, \frac{x^{m_\ell}-\lambda_\ell}{g_\ell\left(x\right)}$ are pairwise coprime polynomials in $\mathbb{F}_q[x],$ Corollary 8 and Theorem 9 of \cite{ramy}  provided methods to construct LCD codes over finite fields. LCD codes over finite fields are primarily studied due to their application in constructing orthogonal direct-sum masking schemes for protecting sensitive data against fault-injection and side-channel attacks. The security threshold (against fault-injection and side-channel attacks) of the orthogonal direct-sum masking scheme based on an LCD code is equal to the Hamming distance of the LCD code. 
Now, when $\frac{x^{m_1}-\lambda_1}{g_1\left(x\right)}, \frac{x^{m_2}-\lambda_2}{g_2\left(x\right)}, \ldots, \frac{x^{m_\ell}-\lambda_\ell}{g_\ell\left(x\right)}$ are pairwise coprime polynomials in $\mathbb{F}_q[x],$ by Theorem  7 of \cite{ramy}, it follows that the code $\mathcal{C}$ (as defined in Theorem 7 of \cite{ramy}) is LCD if and only if the $\lambda_i$-constacyclic code $\pi_i(\mathcal{C})$ is LCD for $1 \leq i \leq \ell.$ Thus the security thresholds $S,$ $S_1,S_2,\ldots,S_{\ell}$ against fault-injection and side-channel attacks of the orthogonal direct-sum masking schemes constructed using the LCD codes $\mathcal{C},$ $\pi_1(\mathcal{C}),\pi_2(\mathcal{C}),\ldots, \pi_{\ell}(\mathcal{C})$ respectively, satisfy the following:
$$S=d(\mathcal{C})\text{ and }S_i=d(\pi_i(\mathcal{C}))\text{~ for ~}1 \leq i \leq \ell,$$ which, by \eqref{D}, gives $$S=\min\{S_1,S_2,\cdots,S_{\ell}\}.$$ That is, the security threshold of the orthogonal direct-sum masking scheme designed using the LCD code $\mathcal{C}$ can not exceed the  security threshold of the orthogonal direct-sum masking schemes designed using any of the LCD constacyclic codes $\pi_1(\mathcal{C}),\pi_2(\mathcal{C}),\ldots, \pi_{\ell}(\mathcal{C}).$ 
Therefore, the LCD codes obtained in Corollary 8 and Theorem 9 of \cite{ramy} hold limited practical significance.}

In Corollary 3.1 of  \cite{multi2}, we obtained the dimension of a $\rho$-generator multi-twisted code of block lengths $(m_1,m_2,\ldots, m_\ell)$ over $\mathbb{F}_q$ by summing up the degrees of the diagonal polynomials in the normalized generating set of the code, where $\rho \geq 1$ is any integer and $m_1,m_2,\ldots, m_{\ell}$ are arbitrary positive integers, not necessarily coprime to $q$. In \cite{ramy}, the authors highlighted the computation of a normalized generating set for a multi-twisted code as a drawback of Corollary 3.1 of \cite{multi2} used to determine the dimension of the multi-twisted code (see page 2, lines 27-29 of \cite{ramy}), which is again an unsubstantiated claim, as the proof of Theorem 3.1 of \cite{multi2} outlines an approach to find a normalized generating set of the multi-twisted code. Highlighting this drawback of Corollary 3.1 of \cite{multi2},  the authors, in Theorem 13 of \cite{ramy}, suggested an alternate approach to obtain the dimension of a $\rho$-generator multi-twisted code of block lengths $(m_1,m_2,\ldots, m_\ell)$ over $\mathbb{F}_q$, where one needs to obtain the greatest common divisor of all $\ell \times \ell$ minors of its generator matrix, which is a $(\rho +\ell) \times \ell$ matrix over $\mathbb{F}_q[x].$ We believe that computing $\ell \times \ell$ minors of a generator matrix of a $\rho$-generator multi-twisted code of block lengths $(m_1,m_2,\ldots, m_\ell)$ over $\mathbb{F}_q$ and then finding their greatest common divisor may be computationally hard for large values of $\ell$ and $\rho$ in general.   In \cite{ramy}, the authors also  claim that Theorem 13 of \cite{ramy} generalizes Theorem 5.1(b) of  \cite{multi} (see page 8, the first sentence in Section 4 and page 9, lines 4-5 of \cite{ramy}), which is also not true, as Theorem 13 of \cite{ramy} does not coincide with Theorem 5.1(b) of \cite{multi} when $\rho =1$. 

In our opinion, several claims made by the authors in \cite{ramy} are either factually incorrect or  lack adequate substantiation, which may confuse the readers about the contributions of \cite{multi2, multi}. Our remarks on the work \cite{ramy} intend to provide the clarity and inform about the  true contributions and findings of our research.

\end{document}